\documentclass[conference]{IEEEtran}

\usepackage{epsfig}
\usepackage{latexsym}
\usepackage{amsfonts}
\usepackage{xspace}
\usepackage{amssymb}
\usepackage{psfrag}
\usepackage{subfigure}
\usepackage{url}
\usepackage{epsfig}
\usepackage{epsf}
\usepackage{array}
\usepackage{pstricks}
\usepackage[cmex10]{amsmath}
\usepackage{cite}
\usepackage{listings}

\begin{document}
%
\title{Quantitative Analysis for Authentication of Low-cost RFID Tags}


\author{\IEEEauthorblockN{Ioannis~Paparrizos}
\IEEEauthorblockA{School of Computer and\\ Communication Sciences\\
EPFL, \\Switzerland\\
Email: ioannis.paparrizos@epfl.ch}
\and
\IEEEauthorblockN{Stylianos~Basagiannis}
\IEEEauthorblockA{Department of Informatics\\ Aristotle University of\\
Thessaloniki\\
Greece 54124\\
Email: basags@csd.auth.gr}
\and
\IEEEauthorblockN{Sophia~Petridou}
\IEEEauthorblockA{Department of Informatics\\ Aristotle University of\\
Thessaloniki\\
Greece 54124\\
Email: spetrido@csd.auth.gr}}

\maketitle

\begin{abstract}
Formal analysis techniques are widely used today in order to verify and analyze communication protocols. In this work, we launch a quantitative verification analysis for the low-cost Radio Frequency Identification (RFID) protocol proposed by Song and Mitchell. The analysis exploits a Discrete-Time Markov Chain (DTMC) using the well-known PRISM model checker. We have managed to represent up to $100$ RFID tags communicating with a reader and quantify each RFID session according to the protocol's computation and transmission cost requirements. As a consequence, not only does the proposed analysis provide quantitative verification results, but also it constitutes a methodology for RFID designers who want to validate their products under specific cost requirements.
\end{abstract}


\IEEEpeerreviewmaketitle

\section{Introduction}\label{intro}
\IEEEPARstart{F}{ormal} analysis techniques, such as probabilistic model checking, are widely used today in order to analyze and verify communication protocols~\cite{Kwiatkowska,BasagiannisCOSE}. In bibliography, security protocols being published with flaws~\cite{Lowe96,Lowe97,BasagiannisCOMPSAC} constitute examples that empower the necessity of using formal methods prior to the design and implementation of a communication protocol. At the same time, given that security is a fundamental issue in communication protocols~\cite{BasagiannisCOSE,PetridouBTS}, quantitative formal analysis can be applied to obtain useful results regarding both the validation of their security properties and the cost to support them~\cite{BasagiannisCOSE,PetridouISCC}. This is important, since the tradeoff of gaining in security is losing in terms of computation cost. Therefore, cost should not been overlooked throughout quantitative analysis, since it can be a prohibited design parameter, especially for protocols executed by low-cost hardware devices, such as RFID tags.

RFID tags are used in industry for supply-chain management, payment systems and inventory monitoring~\cite{Rizomiliotis} and constitute one of the three ($3$) basic entities of an RFID system along with RFID readers and a server. One of the great challenges in the field of RFID is the integration of secure tag identification with low-cost computation and memory expenditure~\cite{SongM}. This requirement forces the tag manufacturers to look for lightweight authentication solutions which preserve security guaranties of an RFID protocol session. In a real-world scenario, RFID protocol operates in a multi-parallel session environment where a large number of sessions between tags and reader will be established concurrently. The latter rises up questions about the overall computation and transmission cost for a reader-server to identify a group of tags.

In this work, we propose the use of probabilistic model checking~\cite{KwiatkowskaQUEST} to verify the Song and Mitchell's RFID authentication protocol~\cite{SongM}. We develop a Discrete-Time Markov Chain (DTMC) model~\cite{Hansson} which represents a multiple tag RFID scheme where tags' authentications are validated. In the PRISM framework, the aforementioned DTMC model is augmented with computation and transmission cost requirements derived by~\cite{SongM}. We produce quantitative results for computation cost of server and tags and for transmission cost regarding up to $100$ simultaneous parallel sessions. We also provide server processing time and tags' time delay results. To the best of our knowledge, this is the first research effort that performs a quantitative analysis of an RFID protocol using probabilistic model checking.

\section{Song-Mitchell's RFID Protocol}\label{smrfid}
The Song and Mitchell's protocol is a well-known authentication protocol for low-cost RFID tags~\cite{SongM}. It comprises three ($3$) basic entities, namely, a group of RFID tags $T_i$, an RFID reader $R$ that radio-communicates with $T_i$ and a back-end server $S$ that contains the record identification database for each tag $T_i$, i.e. $[(u_i,t_i)_{new},(u_i,t_i)_{old},D_i]$. A single protocol session consists of six ($6$) steps, as shown in Fig.~\ref{rfid}. According to the notation provided in Table~\ref{notation} these steps are summarized as follows:
\begin{figure*}[!t]
{\centering \includegraphics[scale=0.66] {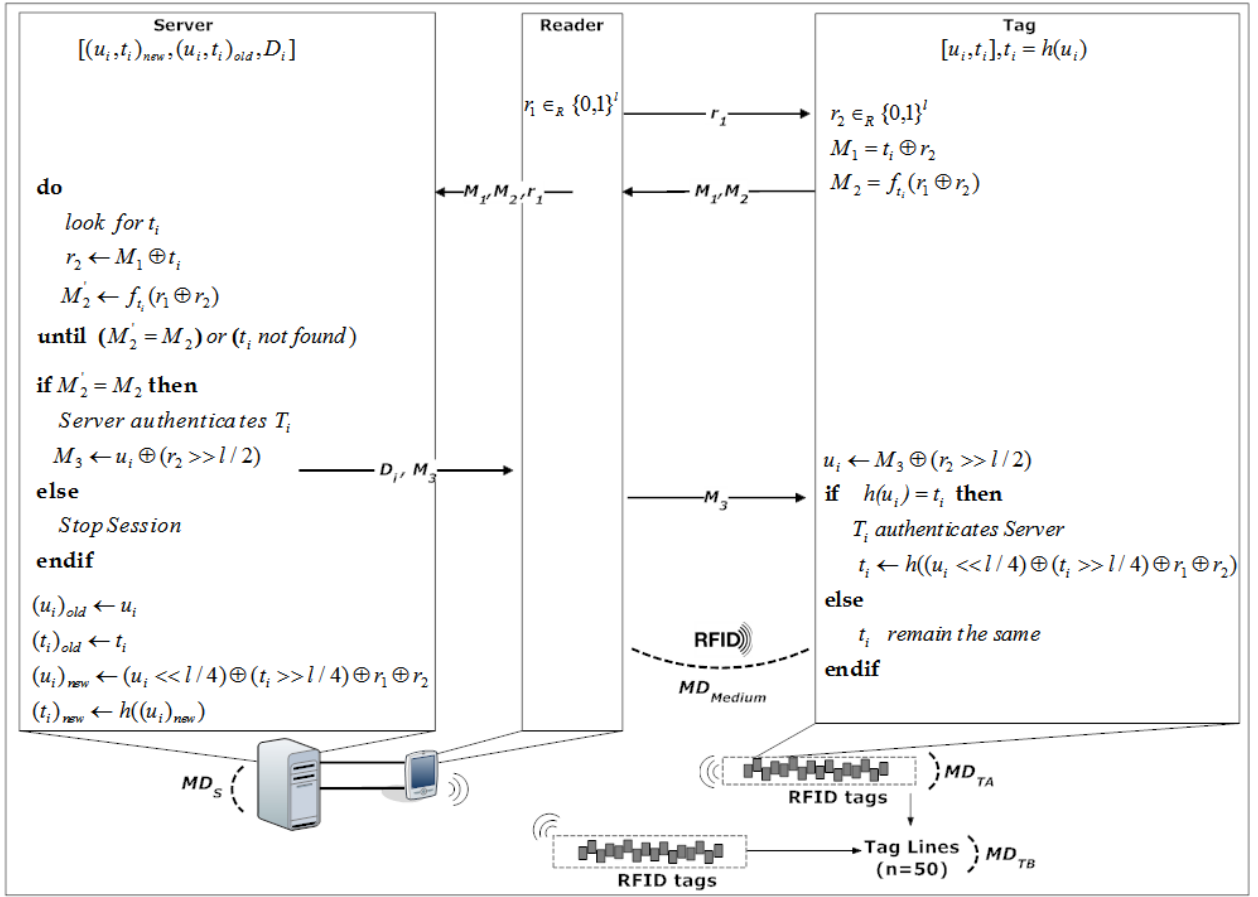} \caption{The analyzed Song-Mitchell's RFID authentication protocol}\label{rfid}}
\end{figure*}
\begin{enumerate}
  \item Reader $R$ generates a random value $r_1\in \Re [0,1]^{l}$ and sends it to $T_i$.
  \item Once $T_i$ receives $r_1$, it generates a random value $r_2\in \Re [0,1]^{l}$, computes $M_1=t_i\oplus r_2$ and $M_2= f_{t_i}(r_1\oplus r_2)$ and sends them to the reader $R$.
  \item $R$ forwards $M_1$, $M_2$ and the random bit-string $r_1$ to the server $S$.
  \item $S$ looks into its tag identity pairs database - both new and old - for a $t_i$ such that $r_2\leftarrow M_1\oplus t_i$ and $M^{'}_2=f_{t_i}(r_1\oplus r_2)=M_2$. If no suitable $t_i$ is found, $S$ sends an error message to $R$ and stops the session. Otherwise, $T_i$ has been authenticated by $S$ which, in turn, computes $M_3=u_i\oplus (r_2 \gg l /2)$ and sends it to $R$ along with $D_i$. After $M_3$ transmission, $S$ updates its tag database as follows: sets $u_{i(old)}$, $t_{i(old)}$ to $u_i$ and $t_i$, respectively, and $u_{i(new)}$, $t_{i(new)}$ to $(u_i\ll l/4)\oplus(t_i\gg l/4)\oplus r_1 \oplus r_2$ and $h(u_{i(new)})$, respectively.
  \item $R$ forwards $M_3$ to $T_i$.
  \item Upon receipt of $M_3$, $T_i$ computes $u_i \leftarrow M_3\oplus(r2\gg l/2)$ and checks if $h(u_i)=t_i$. If the check is true, then $S$ has been authenticated by $T_i$ and $T_i$ updates $t_i$ to $h((u_i\ll l/4)\oplus(t_i\gg l/4)\oplus r_1 \oplus r_2)$. Otherwise, $t_i$ remains the same.
\end{enumerate}

\begin{table}
\caption{Table of Notation}\label{notation}
\begin{tabular}{p{3.5cm} p{4.5cm}}
    \hline   Symbol & Description
\\  \hline $T=\{T_{1},\ldots,T_{n}\}$ & Group of tags $T$, $i=1,\ldots,n$
\\
    $n$ & Number of tags, $n=1,\ldots,50$
\\
    $R$ & RFID reader
\\
    $S$ & Back-end server
\\
    $h$ & Hash function
\\
    $f_k$ & Keyed hash function
\\
    $l$ & The bit-length of a tag identifier
\\
    $D_i$ & Information associated with tag $T_i$
\\
    $u_i$ & An $l$-bit string assigned to $T_i$
\\
    $t_i$ & $T_i$'s $l$-bit Identifier, $t_i=h(u_i)$
\\
    $x_{new}$ & The updated value of $x$
\\
    $x_{old}$ & The most recent value of $x$
\\
    $r$ & Random string of $l$ bits
\\
    $\oplus$ & $XOR$ operator
\\  \hline
\end{tabular}
\end{table}

In the above communication, the channel between the server $S$ and the reader $R$ is secure, while $R$ and $T_i$ communicate over an insecure channel. The proposed model considers two different groups of tags, namely the groups $T_A$ and $T_B$, with $n=1,\ldots,50$ tags $T_i$ per group. Given this range of $n$, we define $N=2,\ldots,100$ to be the upper bound of tags that the server $S$ can authenticate concurrently.

\section{RFID modeling Using DTMC}\label{principles}
The proposed analysis is based on probabilistic model checking principles. The RFID protocol to be analyzed is modeled using DTMCs in the PRISM model checking framework. The proposed model is augmented with computation and transmission cost requirements derived by~\cite{SongM}. The reader may obtain the developed RFID-DTMC model from~\cite{RFIDDTMC}.

In PRISM, a probabilistic model is defined as a set of $m$ modules , $MD=\{MD_1,\ldots,MD_m\}$. Each $MD_i$ module is defined as a pair of $(Var_i,C_i)$, where $Var_i$ is a set of integer-valued local variables with finite range and $C_i$ is a set of commands. The set $Var_i$ defines the local state space of module $MD_i$ and in turn $Var$ denotes the set of all local variables of the model, i.e., $Var=\bigcup^{i=1}_{m}Var_i$. Furthermore, each variable $v \in Var$ has an initial value $\bar{v}$.

Our DTMC model includes $m=4$ modules, namely, $MD_S$ representing both the server $S$ and the reader $R$, $MD_{T_A}$ and $MD_{T_B}$ for groups of tags $T_A$ and $T_B$ and $MD_{Medium}$ for the communication medium between $MD_S$, $MD_{T_A}$ and $MD_{T_B}$. The behavior of a module $MD_i$ is defined by the set of commands $C_i$. Each command $c \in C_i$ takes the form of $(g, (\lambda_1, u_1),\ldots,(\lambda_{n_c},u_{n_c}))$, comprising a guard $g$ and a set of pairs $(\lambda_j,u_j)$, where $\lambda_j \in \Re_{>0}$ and $u_j$ is an update for each $1 \leq j\leq n_c$. A guard $g$ is a predicate over the set of all local variables $Var$ and each $u_j$ update corresponds to a possible transition of module $MD_i$. If $Var_i$ contains $n_i$ local variables, $\{v_1,\ldots,v_{n_i}\}$, then an update takes the form $(v_1'=expr_1)\cap \ldots \cap(v_{n_i}'=expr_{n_i})$, where $expr_j$ is an expression in terms of the variables in $Var$. Information of the model may be omitted if an update $u_i$ does not affect some variables $Var_i$. In DTMC model specification, the constants $\lambda_j$ determines the probability attached to transitions (i.e., the probability attached to transition that the update takes place), thus, $\lambda_j\in (0,1]$ for $1 \leq j\leq n_c$ and $\sum^{n_c}_{j=1}\lambda_j=1$~\cite{Bianco}.

More specifically, a DTMC model is defined as a tuple $(S, \bar{s}, P, L )$, where:
\begin{itemize}
\item $S$ is a finite set of states
\item $\bar{s} \in S$ is the initial state
\item $P: S \times S \rightarrow \Re_{\geq 0}$ is the transition probability matrix such that $\sum_{s'\in S} P(s,s')=1$ and
\item $L: S \rightarrow 2^{AP}$ is a labeling function mapping states to sets of atomic propositions from a set $AP$ with the properties of interest
\end{itemize}

Terminating states are modeled by a single transition going back to the same state with probability $1$. DTMCs are further described in~\cite{Hansson}. In order to attach RFID cost parameters into the developed DTMC model, we define reward modules $MD_{RC}$ and $MD_{RT}$ which correspond to the RFID computation and transmission requirements, respectively~\cite{SongM}.

Results will be acquired by defining the appropriate formulae properties according to the Probabilistic Computational Tree Logic (PCTL)~\cite{KwiatkowskaSFM}. The syntax of PCTL is as follows:
    \begin{displaymath}
    \phi\ ::=\ true\ |\ \alpha\ |\ \phi\ \wedge\ \phi\ |\ \neg\phi\ |\ P_{\bowtie p}[\psi]
    \end{displaymath}
    \begin{displaymath}
    \psi\ ::=\ X\ \phi\ |\ \phi\ \mathcal{U}^{\leq t}\ \phi\ |\ \phi\ \mathcal{U}\ \phi
    \end{displaymath}
where $a$ is an atomic proposition, operator $\bowtie \in [\leq,<,\geq,>]$, $p\in[0,1]$ and $t\in \aleph_{\geq 0}$.

For a DTMC $(S, \bar{s}, P, L )$, a reward structure is a tuple ($\varrho, \iota$), where:
 \begin{itemize}
   \item $\varrho:S\rightarrow\Re_{\geq 0}$ is a vector of state rewards, and
   \item $\iota:S\times S\rightarrow\Re_{\geq 0}$ is a matrix of transition rewards.
 \end{itemize}
DTMC allows the specification of four distinct types of rewards $R$:
\begin{itemize}
  \item \emph{Instantaneous} $R_{\bowtie r}[\mathrm{I}^{=t}]$: the expected value of the reward at time-instant $t$ is $\bowtie r$,
  \item \emph{Cumulative} $R_{\bowtie r}[\mathrm{C}^{\leq t}]$: the expected reward cumulated up to time-instant $t$ is $\bowtie r$,
  \item \emph{Reachability} $R_{\bowtie r}[\mathrm{F}\ \phi]$: the expected reward cumulated before reaching $\phi$ is $\bowtie r$,
  \item \emph{Steady-state} $R_{\bowtie r}[\mathcal{S}]$: the long-run average expected reward is $\bowtie r$.
\end{itemize}
Cumulative and rechability reward properties are employed for the proposed quantitative verification of the proposed RFID-DTMC model.

\section{Quantitative Verification Results}\label{results}
\begin{figure}
{\centering \includegraphics[scale=0.43] {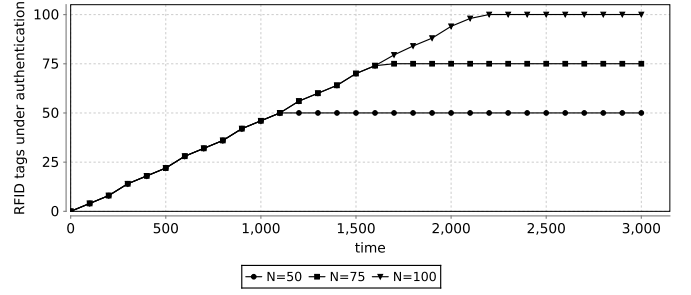} \caption{The number of RFID tags under authentication as a function of time for different upper bound of tags $N$}\label{r1}}
\end{figure}

\begin{figure}
{\centering \includegraphics[scale=0.43] {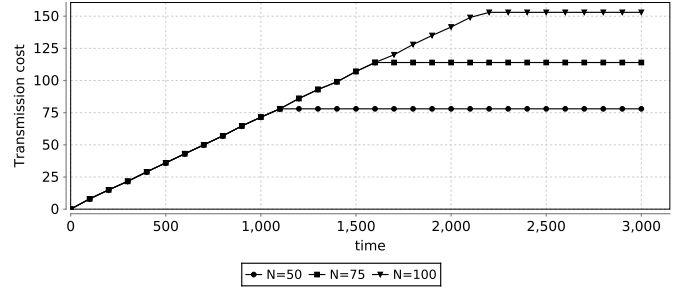} \caption{Transmission cost of RFID authentication protocol as a function of time for different upper bound of tags $N$}\label{r2}}
\end{figure}

The novelty of the current work is that for the first time probabilistic model checking using DTMCs is employed in order to verify the properties of the RFID authentication protocol. In the proposed quantitative analysis we model multiple RFID sessions according to the steps described in Section~\ref{smrfid}. Since, our model considers two groups $T_A$ and $T_B$ of up to $50$ tags each, it concludes that the server $S$ can authenticate concurrently an upper bound of up to $N=100$ tags.

Fig.~\ref{r1} represents the number of RFID tags which are under authentication as a function of time expressed in time steps. It is natural that as time passes the server $S$ will authenticate an increasing number of incoming tags $T_i$. However, this number has a threshold equal to $N$, i.e., the upper bound of tags being authenticated concurrently. We observe that for $N=50, 75, 100$ the corresponding curve is fixed at $N$ and this happens at time step $1100, 1600, 2200$ respectively. This means the smaller the $N$ the sooner the curves' fixing.

In line with the above observation, it is expected that the transmission cost of the RFID protocol, which depends on the number of tags under concurrent authentication, will be increased with time but it will not exceed an up limit which indicates the time that the server $S$ is constantly fully occupied with $N$ tags. Thus, Fig.~\ref{r2} confirms that at time step $1100, 1600, 2200$ the tags' authentication requests will be fixed at their maximum keeping transmission cost unchanged. Results depicted in Fig.~\ref{r1} and \ref{r2} derived using cumulative reward queries.

\begin{figure}
{\centering \includegraphics[scale=0.43] {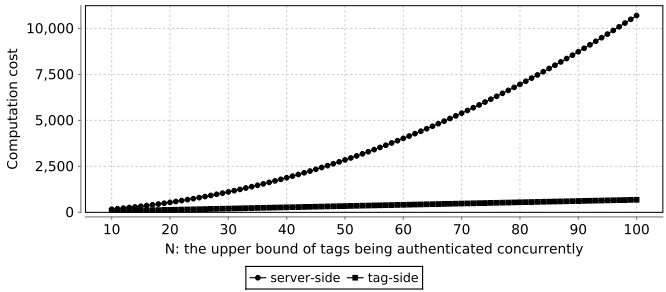} \caption{Server- and tag-side computation cost as a function of the upper bound of tags $N$}\label{r3}}
\end{figure}

Apart from transmission cost, the proposed analysis incorporates the computation requirements of RFID protocol. More specifically, according to~\cite{SongM}, the server-side and tag-side cost is exponential and linear to the number of tags, respectively. In Fig.~\ref{r3} we depict computation cost at server- and tag-side as a function of $N$, for $N=10,\ldots,100$, and we confirm the expected curves' trend.

\begin{figure}
{\centering \includegraphics[scale=0.43] {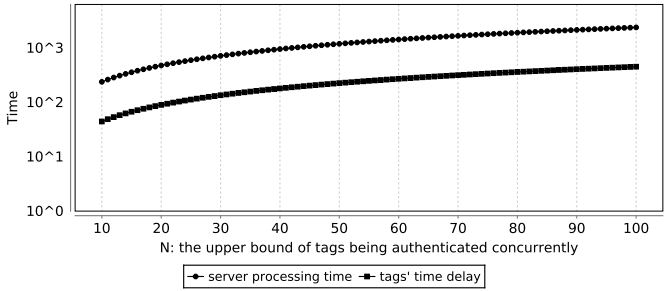} \caption{Server processing time and tags' time delay as a function of the upper bound of tags $N$}\label{r4}}
\end{figure}

We finally launch a set of experiments in order to compute the service rate and mean tags' delay. Fig.~\ref{r4} shows that both the server processing time and tags' time delay are increased in line with $N$, for $N=10,\ldots,100$. Service rate is equal to $25$ tags per time step while mean tags' delay is approximately $4.5$ time steps. Results depicted in Fig.~\ref{r3} and \ref{r4} derived using rechability reward queries.

An additional value of the proposed model, besides the above results, is that it is designed to be configurable. Cost requirements incorporating in the proposed model are protocol dependant providing an analyst the capability of assigning rewards according to the hardware specifications of a protocol.

\section{Conclusion}
Quantitative analysis using probabilistic model checking is firstly used in this work in order to verify cost requirements of the Song and Mitchell's RFID authentication protocol, while its security properties are preserved. We have managed to create a representative cost weighted DTMC model within the PRISM model checking environment, towards the quantitative analysis of a parallel session scenario that include up to $100$ RFID tag identifications.

Apart from results launched by the proposed analysis, current work provides insights for addressing cost-related issues of RFID protocols and deciding upon their cost-dependent viability in line with their security guarantees.

Our future plans involve the cost-based analysis of RFID solutions~\cite{Rizomiliotis} which propose some fixes for strengthening the security of RFID protocols. In this way we will be able to evaluate the computation cost caused by a fix solution. Furthermore, our goal is to model and compare a series of Radio Identification protocols using the proposed analysis. In this way we will provide researchers and protocol designers with a complete framework for quantitative analyzing any security mechanism embedded in existing or new RFID protocols, especially when exploiting low-cost hardware.

\bibliographystyle{./IEEEtranS} 
\bibliography{./IEEEabrv,./references}

\end{document}